\documentclass[aps,prl,twocolumn,showpacs,superscriptaddress,superscriptaddress]{revtex4-1}  
\usepackage{graphicx,color}  
\usepackage{dcolumn}   
\usepackage{bm}        
\usepackage{amssymb}   

\usepackage{ulem}
\usepackage[section]{placeins}
\usepackage{float}
\usepackage{amsmath}
\usepackage{epstopdf}

\begin{document}

\title{Evidence for marginal stability in emulsions}

\author{Jie Lin}
\affiliation{Center for Soft Matter Research, Department of Physics, New York University, New York, NY 10003}
\author{ Ivane Jorjadze}
\affiliation{Center for Soft Matter Research, Department of Physics, New York University, New York, NY 10003}
\author{Lea-Laetitia Pontani}
\affiliation{Center for Soft Matter Research, Department of Physics, New York University, New York, NY 10003}
\affiliation{Institut des Nanosciences de Paris, UMR 7588 - CNRS/Universite Pierre et Marie Curie, 75005 Paris, France}
\author{Matthieu Wyart}
\affiliation{Physics Institute, Ecole Polytechnique Federale de Lausanne, 1015 Lausanne, Switzerland}
\affiliation{Center for Soft Matter Research, Department of Physics, New York University, New York, NY 10003}
\author{Jasna Brujic}
\affiliation{Center for Soft Matter Research, Department of Physics, New York University, New York, NY 10003}

\date{\today}

\begin{abstract}
We report the first measurements of the effect of pressure on vibrational modes in emulsions, which serve as a model for soft frictionless spheres at zero temperature. As a function of the applied pressure, we find that the density of states $D(\omega)$ exhibits a low-frequency cutoff $\omega^{\ast}$, which scales linearly with the number of extra contacts per particle $\delta z$. Moreover, for $\omega<\omega^*$, $D(\omega)\sim \omega^2/\omega^{\ast2}$; a quadratic behavior whose prefactor is larger than what is expected from Debye theory. This surprising result agrees with recent theoretical findings \cite{DeGiuli14,Franz15}. Finally, the degree of localization of the softest low frequency modes increases with compression, as shown by the participation ratio as well as their spatial configurations. Overall, our observations show that emulsions are marginally stable and display non-plane-wave modes up to vanishing frequencies.
\end{abstract}

\maketitle

Due to their translational symmetry, the vibrational modes of crystals are plane waves that exhibit the Debye scaling in the density of states, $D(\omega) \sim \omega^{d-1}$, where $d$ is the spatial dimension. By contrast, amorphous materials, such as granular materials and glasses, exhibit peculiar elastic properties. In particular, they display an excess of low frequency vibrational modes in the density of states compared to crystalline solids, known as the ``boson'' peak~\cite{Phillips81, bosp1995, ghosh2010}. In simple model systems, such as sphere packings, the boson peak diverges in amplitude as the pressure is lowered towards the unjamming transition~\cite{ohern2003, wyart2005, silbert2005, silbert2009, xu2010,DeGiuli14b, wyartT2,wyart2005,DeGiuli14, hecke2010}. The associated soft modes are responsible for anomalous elastic and transport properties~\cite{wyartT2,wyart2005,DeGiuli14,Vitelli10,Xu09} and play an important role in plasticity and activated events \cite{manning2011, Chen11, Brito09}. As the applied pressure increases above the jamming transition, their spatial localization increases~\cite{ohern2003, wyart2005, silbert2005, silbert2009, xu2010}. 

Theoretically, it was argued  that these properties result from the fact that packings near jamming are {\it marginally stable}: their structure is such that they lie at the threshold of a linear elastic instability controlled by pressure and coordination~\cite{wyart2005}. Very recently, a mean-field approximation \cite{DeGiuli14} and exact calculations in infinite dimensions~\cite{Franz15} indicated that for marginally stable structures at low frequencies $D(\omega)\sim \omega^2/\omega^{\ast2}$ in any spatial dimension, where $\omega^*$ is a characteristic frequency of the soft modes. In three dimensions, this quadratic behavior matches the Debye result, but with a much larger prefactor. This result remains untested in real materials. 

Experimentally, the density of states of colloidal glasses in 2D has been obtained from the thermal motion of the particles as a function of the global density below the glass transition~\cite{yodh2010, ghosh2010, Chen11, tan2012}. However, experimental tests on athermal particulate packings and in 3D are lacking in the literature, and are appropriate to accurately test if materials are indeed marginally stable. In this letter, we experimentally investigate the pressure dependence of the density of states and the localization of vibrational modes in compressed emulsions. Because emulsions are effectively at zero temperature, the present experiment appears close enough to the jamming transition to precisely probe its properties. We find that the density of states can be rescaled by some frequency $\omega^{\ast}\sim (\phi-\phi_c)^{1/2}$ where $\phi$ is the packing fraction and $\phi_c$ its (preparation dependent) value at the jamming threshold, and find that it is consistent with the recently proposed scaling behavior $D(\omega)\sim (\omega/\omega^{\ast})^2$\cite{DeGiuli14,Charbonneau15}. Finally, we examine the microscopic configuration of the soft modes by measuring the participation ratio of the particles and their spatial distribution. We find delocalized soft modes near jamming, and localized soft modes far above jamming.

\begin{figure}[b!]
\includegraphics[scale=0.6]{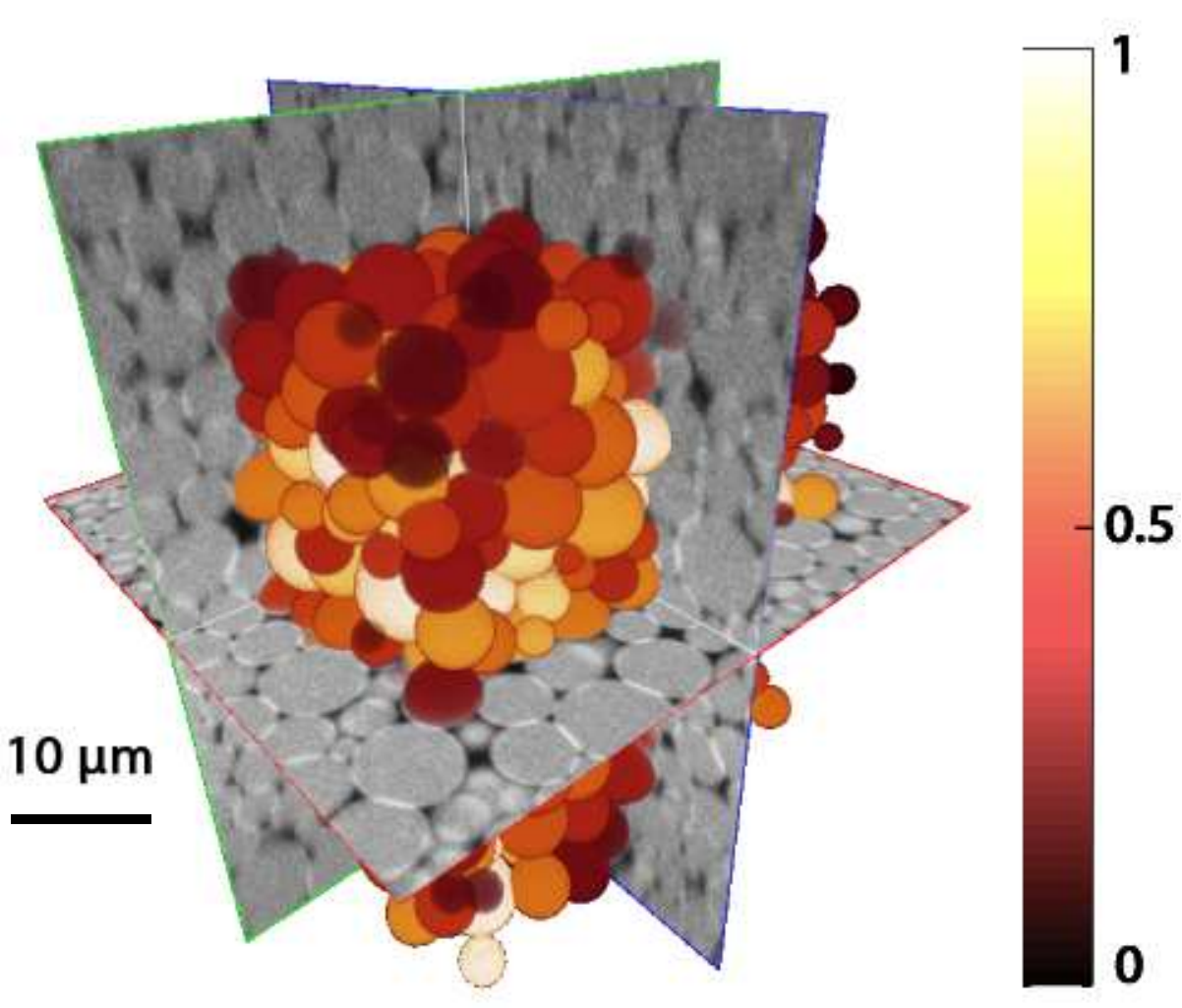}
\caption{\label{fig1} (Color online) Confocal image of an emulsion at a pressure of $0.4$~kPa and $\phi=0.74\pm 0.02$ (greyscale). For each reconstructed droplet $i$ in the middle, the low-frequency vibrational amplitude in the color map (dim to bright) is obtained as $X_{i}(m) = \sum_{n<m} |\mathbf{e}^n_{i}|$, where $\boldsymbol{e}^n_{i}$ is the polarizaton vector of the $n$th eigenmode, summing over the lowest $m=40$ modes.}
\end{figure}
\begin{figure*}[hbt!]
\includegraphics[scale=1]{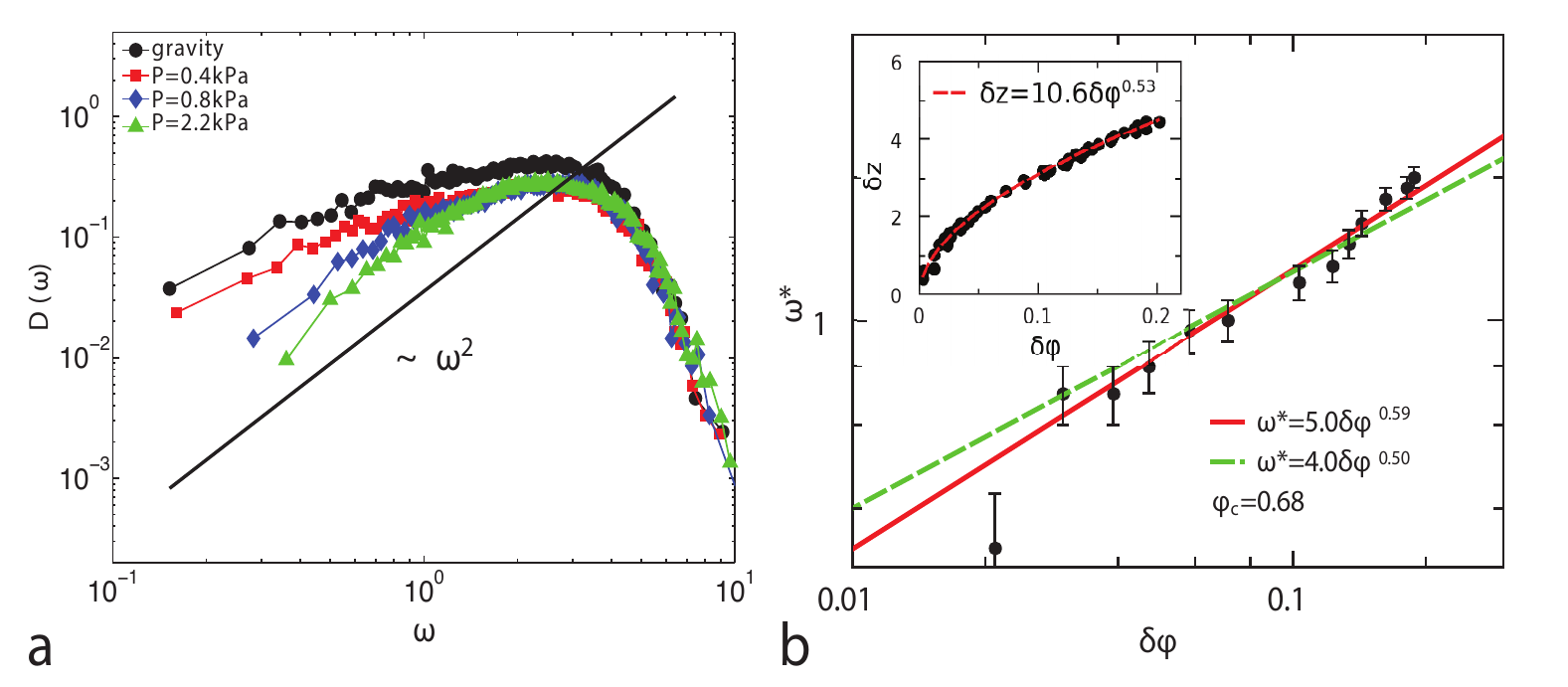}
\caption{\label{fig2} (Color online) (a) The density of states $D(\omega)$ shows a decrease in the number of low frequency modes as the applied pressure is increased. The black line has a slope $2$ in the log-log plot, which is the Debye scaling. (b) The crossover frequency $\omega^{*}$ increases with the distance from jamming $\delta{\phi}=\phi-\phi_c$. The line of best fit (solid line) is in good agreement with the marginal rigidity prediction (dashed line). The inset shows the scaling relation between the excess number of contacts $\delta z$ and $\delta\phi$ to give $\omega^{\ast}\sim \delta z$.}
\end{figure*}

Our experimental system is an oil-in-water emulsion, described in references~\cite{mason1995, jasna2003}, which is a model for soft, frictionless, athermal sphere packings with an average droplet radius $\langle R \rangle= 2.5 \text{ } \mu \text{m}$ and a polydispersity of $25\%$. The refractive index matched emulsion is transparent and the Nile Red dye reveals the 3D packing of droplets using a confocal microscope (Leica TCS SP5 II). Due to a density difference between the droplets and the continuous phase, the lowest compression rate is achieved by creaming under gravity, which gives a packing fraction of $\phi_{\text{c}} = 0.68\pm 0.02$ due to polydispersity, while centrifugation for 20 minutes at an acceleration rate of 3000g leads to highly compressed structures with $\phi=0.88\pm 0.02$ \cite{Jorjadze12a}. Allowing the emulsion to relax to its uncompressed state over a period of several days probes a broad range of intermediate packing densities~\cite{Jorjadze12a}, an example of which is shown in Fig.~\ref{fig1}. The packings are analyzed using a Fourier transform algorithm to identify the particle positions and radii~\cite{jasna2003}. We then extract the repulsive force $\boldsymbol{f}_{\!ij} (=-\boldsymbol{f}_{\!ji})$ exerted  by particle $j$ on particle $i$ via the Princen model~\cite{princ1983}:
\begin{equation}
\boldsymbol f_{\!ij}=-\frac{\sigma}{\widetilde{R}_{ij} A_{ij}} \boldsymbol n_{ij}, \qquad \boldsymbol n_{ij} = \frac{\boldsymbol r_i - \boldsymbol r_j}{|\boldsymbol r_i - \boldsymbol r_j|},
\label{eq1} 
\end{equation}
where $\boldsymbol r_i$ and $\boldsymbol r_j$ denote the particle positions, $A_{ij}$ is the area of droplet deformation, or the geometric area of overlap between the reconstructed spheres~\cite{jasna2007}, $\widetilde{R}_{ij}=\frac{2R_{i}R_{j}}{R_{i}+R_{j}}$ is the weighted mean radius of the droplets, and $\sigma = 9.2 \text{ } \text{mN/m}$ is the interfacial tension. We therefore sum over the force moment tensor for all contacts $\langle ij\rangle$ and then divide by the volume of the box to obtain the total pressure on the system,
\begin{equation}
P=\frac{1}{3V} \sum_{i,j} \boldsymbol f_{\!ij}\cdot  \boldsymbol r_{ij}, \qquad \boldsymbol r_{ij}=\boldsymbol r_j-\boldsymbol r_i
\end{equation}
From the forces, we construct the interaction potential between contacting particles $U(r_{ij})$, where $r_{ij}=|\boldsymbol r_{ij}|$. This potential is harmonic for small deformations. To ensure mechanical equilibrium, we numerically quench the system into the nearest energy minimum to satisfy force balance using the method of steepest descent~\cite{ohern2003, silbert2005}. This procedure does not move any of the particles beyond the resolution of the droplet finding technique, which is one third of a voxel size (i.e. $100$nm, or $2\%$ of the average droplet diameter). This correction does not have a measurable effect on the calculation of the density of states done below and gives validity to the force model in Eq.~\eqref{eq1}. 

We then calculate the corresponding Hessian matrices, defined for every pair $i$, $j$ of particles in contact as $H_{ij}=H^0_{ij}+ H^1_{ij}$, where the first term stems from the contact stiffness and the second from the contact force:
\begin{equation}
H^0_{ij}=\delta_{ij} {\sum_{k\sim i}} \boldsymbol n_{ik} \boldsymbol n_{ik}^T \frac{d^2U}{dr_{ik}^2}-\ \boldsymbol n_{ij} \boldsymbol n_{ij}^T \frac{d^2U}{dr_{ij}^2}\label{H0}
\end{equation}
and
\begin{equation}
\begin{aligned}
H^1_{ij} =\delta_{ij} {\sum_{k\sim i}} \frac{(\text{I}-\boldsymbol n_{ik} \boldsymbol n_{ik}^T)}{r_{ik}}\frac{dU}{dr_{ik}}-\ \frac{(\text{I}-\boldsymbol n_{ij}\boldsymbol  n_{ij}^T )}{r_{ij}}\frac{dU}{dr_{ij}}
\end{aligned}\label{H1}
\end{equation}
where the notation ${k\sim i}$ indicates that particle  $k$ is in contact with particle $i$.
The polarization vectors $\boldsymbol e_i^n$ and the eigenfrequencies $\omega_n$ can now be calculated via  solution of the eigenvalue problem
\begin{equation}
\omega^2_n m_i \boldsymbol e_i^n = \sum_{j} H_{ij} \boldsymbol e_j^n
\label{007}
\end{equation}
where $m_i$ denotes the mass of particle $i$. Note that in our system inertia is negligible, and dynamically modes would be overdamped. Here we have included the masses to allow for a closer comparison with the literature. An alternative choice 
is to set $m_i=1$ for all $i$. Then our analysis would correspond to the normal modes of the hessian, which characterize the amplitude of thermal fluctuations (independently of the damping mechanism). We have checked that both choices lead to essentially identical results.

The amplitudes of the polarization vectors $\boldsymbol e_i^n$ are visualized in Fig.~\ref{fig1}. Since each particle must have all its neighbors and forces identified, we use $\approx1000$ particles in the middle of the experimental packing for each compression and repeat the experiment three times to collect statistics. We thus obtain a histogram of eigenmodes, i.e. the density of states $D(\omega)$,  where $\omega$ is shown in units of natural frequency in Fig.~\ref{fig2}(a) for different levels of compression. The microscopic frequency is given by $\sqrt{\kappa/m}=7.5*10^5$rad/s, where stiffness $\kappa=\sigma/\langle R \rangle$.

As expected, the log-log representation reveals a cut-off frequency at high values, and a progressive increase in the number of low-frequency modes as we approach the jamming transition under gravity, in agreement with numerical simulations~\cite{ohern2003, wyart2005, silbert2005, silbert2009}. Excess soft modes are generally characterized by the ratio $D(\omega)/\omega^{d-1}$, which often presents a maximum in simulations and in experiments on molecular glasses~\cite{bosp1995, niss2007}. Our results do not reveal a maximum in $D(\omega)/\omega^2$, as shown in the Supplemental Material, similar to previously observed data from numerical packings of soft spheres ~\cite{silbert2005}. 

This absence of a peak can be understood in terms of the effect of the experimental control parameter, the pressure, on the system. Pressure affects vibrational properties in two distinct ways: first, it affects the packing fraction, which in turn increases the mean number of contacts between particles, or coordination $z$, and tends to stabilize the system further. This is illustrated in the inset of Fig.\ref{fig2}(b), from previous measurements in the same set-up \cite{Jorjadze12a}. Second, pressure directly has a destabilizing effect that can lead to a "buckling" phenomenon  where soft modes become unstable~\cite{wyart2005,DeGiuli14}. These effects can be disentangled by considering the density of states with and without the applied stress term, i.e. by using $H=H_0+H_1$ or $H=H_0$,  respectively in Eq.\ref{007}. In the latter case, we keep the geometric configurations that were formed due to the applied pressure, but we effectively replace our droplets by point particles connected with relaxed springs. As predicted~\cite{wyart2005}, we find that $D(\omega)$ is affected by pressure only below a certain crossover frequency, $\omega^{\ast}$, which is extracted as the frequency below which the densities of states with and without the pressure term begin to deviate (See Fig. S1 in SI ). Note that $D(\omega)/\omega^2$ has a maximum near $\omega^{\ast}$ if $H_0$ alone is considered.

Marginal stability~\cite{wyart2005} indicates that as pressure increases, the stabilizing effect of increasing coordination (apparent when considering $H_0$ alone) precisely matches the direct destabilizing effect of the pressure (induced by the term $H_1$ in the hessian). In other words, the packing just makes enough contacts to remain stable. As a result, it can be proven that $\omega^*\sim \delta\phi^{1/2}$ \cite{wyart2005}.  We confirm  this prediction in  Fig.\ref{fig2}(b). These findings agree quantitatively with theory and numerical simulations~\cite{silbert2005}, unlike recent colloidal experiments probing the jamming transition~\cite{yodh2010, ikeda2013}. 

\begin{figure}
\includegraphics[scale=0.3]{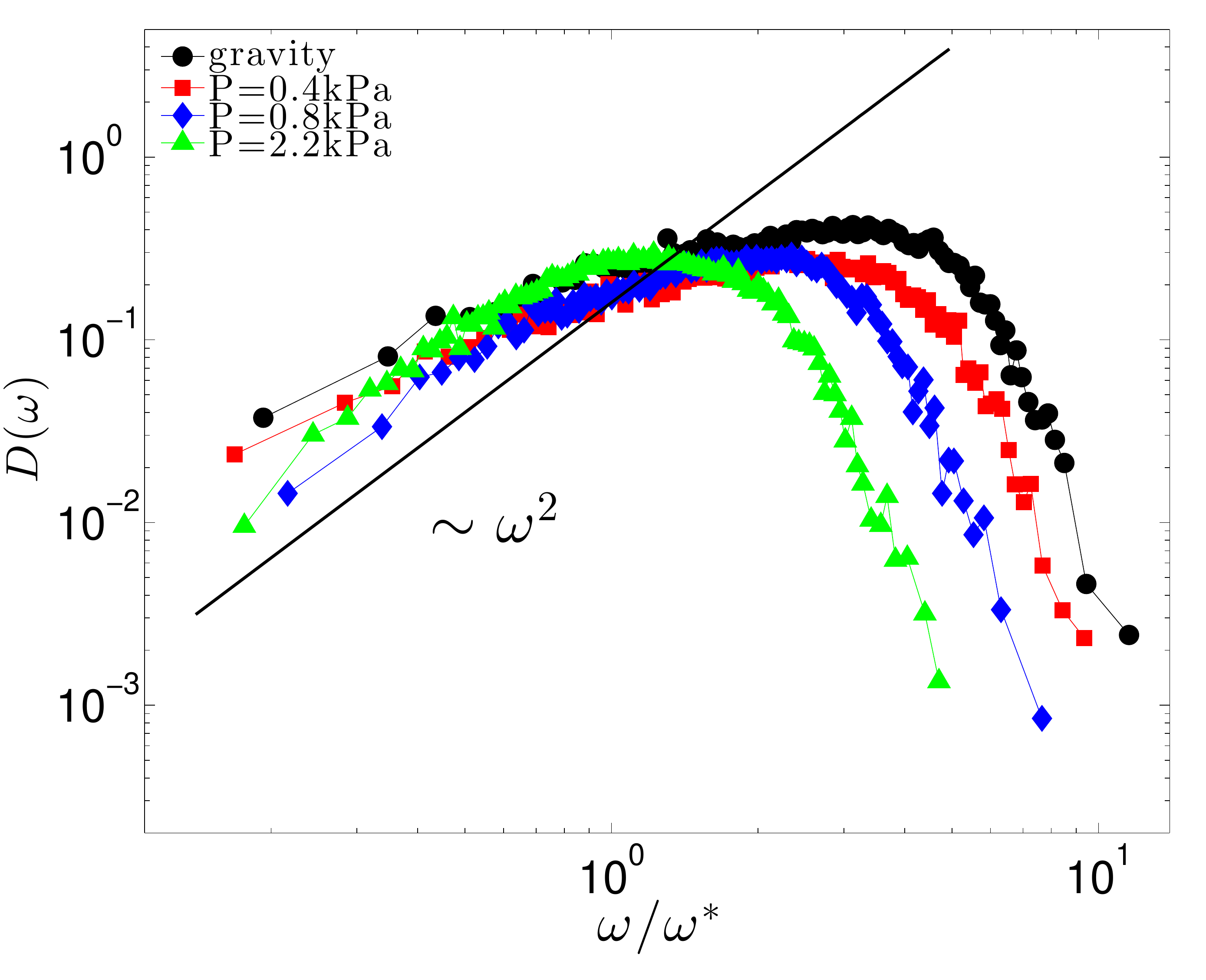}
\caption{The density of states $D(\omega)$ for the rescaled $\omega/\omega^{\ast}$ for all pressures shows good agreement with the predicted slope of $\omega^2$ in Eq.\ref{dw}.} \label{Fig.3} 
\end{figure}

\begin{figure*}[thb!]
\includegraphics[scale=0.7]{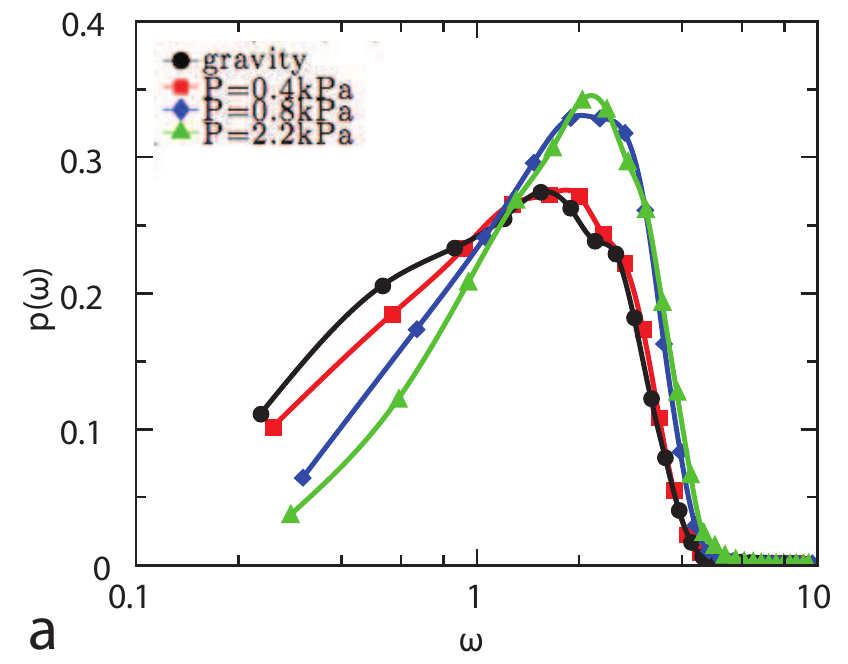}
\includegraphics[scale=0.7]{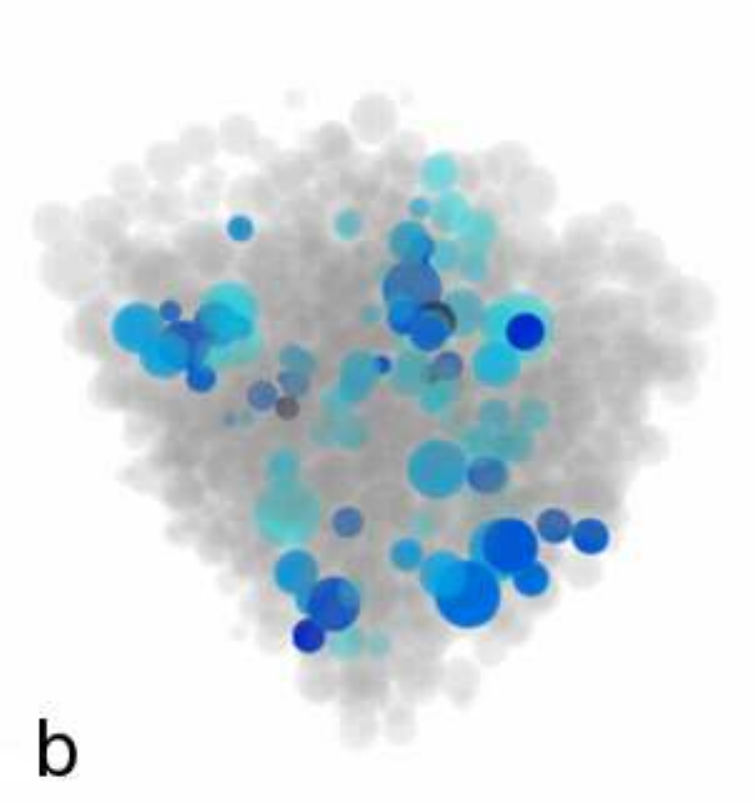}
\includegraphics[scale=0.7]{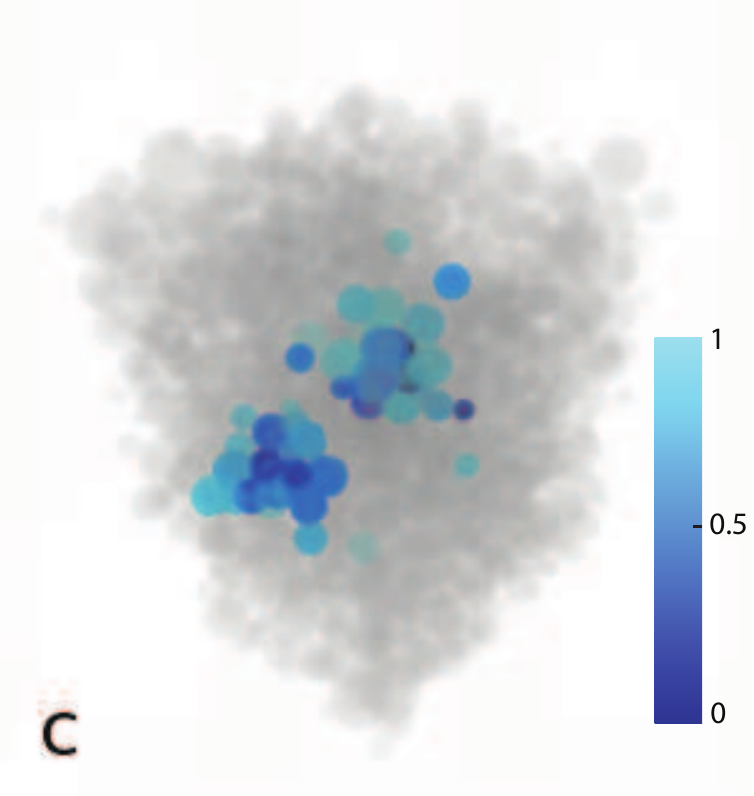}
\caption{\label{fig3} (Color online) (a) Participation ratio, $p(\omega)$. At low frequencies samples at high pressures have lower values of $p(\omega)$, suggesting that these frequencies are more localized. Droplets at low (b) and high (c) pressures colored according to their vibrational amplitudes at the lowest frequency demonstrate visually that an increase in the applied pressure leads to mode localization.}
\end{figure*}

Another prediction of marginal stability is that there is no Debye regime at all.
Instead, the theory of marginal stability leads to
\begin{equation}
D(\omega)\sim (\omega/\omega^{\ast})^{2}\label{dw}
\end{equation}
below $\omega^{\ast}$  \cite{ DeGiuli14, Franz15}, with a factor larger than the theoretical prediction for the Debye model~\cite{wyartT2}, which would predict that $D(\omega)\sim \omega^2/\omega^{\ast3/2}$.  
Rescaling our data in Fig.\ref{fig2}(a) by $\omega/\omega^{\ast}$ reveals a satisfying data collapse for all pressures in Fig.\ref{Fig.3}, consistent with Eq.\ref{dw}. This result is in agreement with recent numerical simulations in high dimensions~\cite{Charbonneau15}.
Since $\omega^{\ast}$ decreases to zero at the jamming transition, the regime below $\omega^{\ast}$ becomes progressively smaller, and 
the experimental resolution may not be high enough to definitively rule out the Debye result. However, together with the softening of modes with compression, our results strongly support marginal stability.

Next, we study the degree of localization of vibrational modes by computing the participation ratio $p(\omega)$ of a mode $n$ as a function of the applied pressure:
\begin{equation}
p\left( \omega_n \right)=\frac{ \left( \sum_{i} m_{i} \left| \boldsymbol{e}^n_{i} \right|^{2} \right)^2 }{N \sum_{i} m_{i}^{2}\left| \boldsymbol{e}^n_{n} \right|^{4} };
\label{eq3} 
\end{equation}
The value of $p(\omega)$ ranges from $0$ to $1$, from modes localized to one particle to extended modes, in which all particles participate. Fig.\ref{fig3}(a) shows that the higher the compression and the lower the frequency, the lower the participation ratio, suggesting increasing localization. This observation is consistent with 2D experiments~\cite{yodh2010}, three dimensional~\cite{xu2010} and higher dimensional simulations~\cite{Charbonneau15}. The participation ratio does not report on the spatial configuration of the participating particles. It is therefore useful to visualize the increase in the degree of localization using a color map (from light to dark blue) for the vibrational amplitudes of the lowest frequency mode in the packing under gravity and that of the highest compression, $P=2.7$kPa, as shown in Fig.\ref{fig3}(b,c), respectively. These configurations clearly show the distinction between extended and localized modes. At intermediate frequencies and pressures, the modes exhibit a coexistence between localized and disordered modes.  

In conclusion, we have characterized experimentally  the effect of pressure on the spectrum of the vibrational modes for compressed emulsions.  Our central results are that both the scaling of the characteristic frequency $\omega^{\ast}$ as well as the behavior of $D(\omega)$ for $\omega<\omega^{\ast}$ are consistent with the notion that emulsion are marginally stable. In the range that can be probed experimentally, our observations agree with recent quantitative mean-field theories describing such a marginal behavior \cite{DeGiuli14, Franz15}, giving empirical support for these views.  In the future, it would be very interesting to study if marginal stability is robust to different system preparations, and if it survives at very large compression~\cite{Urbani2016}.

We thank Eric DeGiuli and Eric Vanden-Eijnden 
for insightful discussions. This work was supported primarily by the Materials Research Science and Engineering Center (MRSEC) program of the National Science Foundation under Award Number DMR-1420073.

\section{Supplementary Information}
\section{Extraction of $\omega^{\ast}$}
\begin{figure}
\includegraphics[scale=0.5]{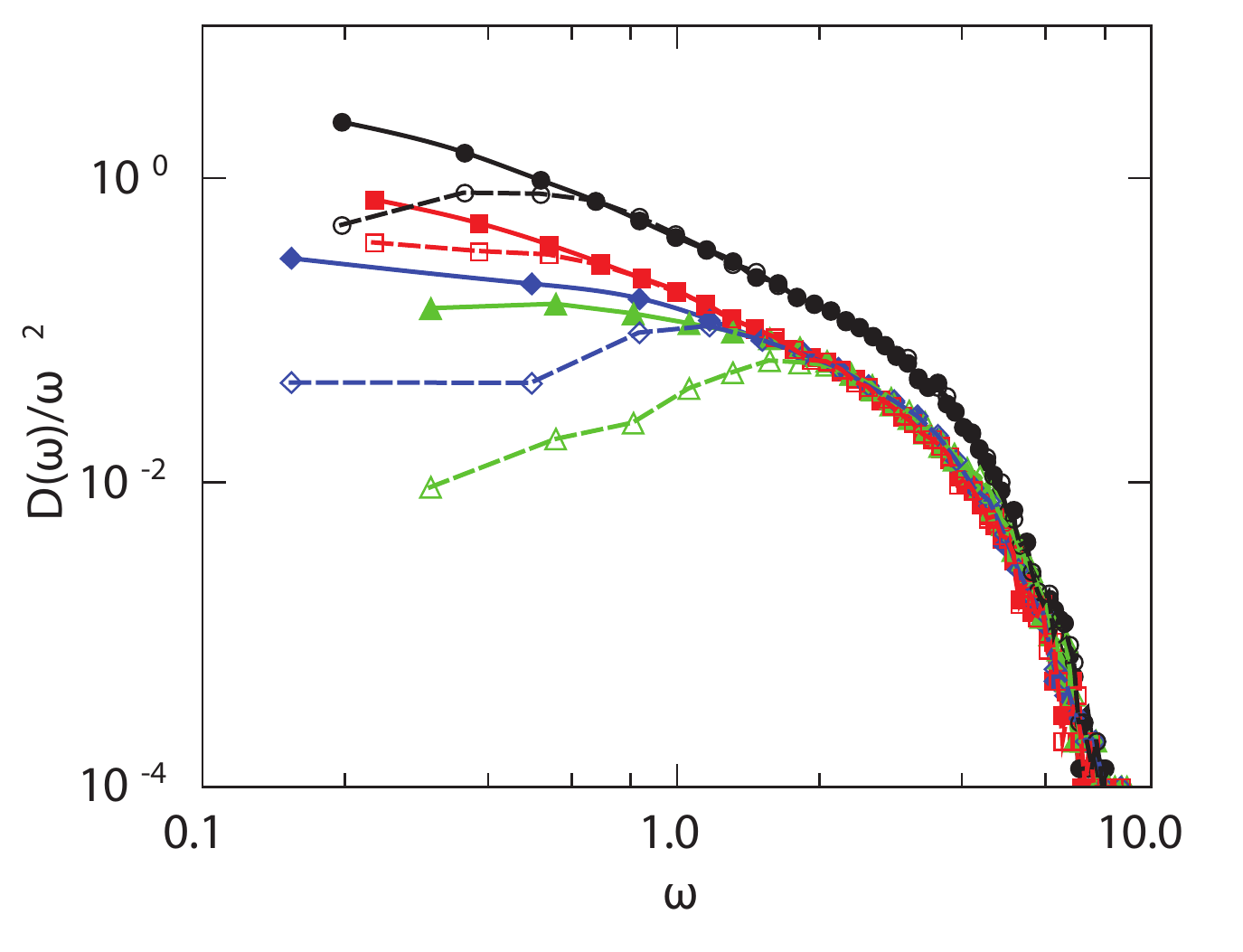}
\caption{\label{SI} (a) The density of states $D(\omega)/\omega^2$ at 4 different pressures. The cutoff $\omega^{*}$ is identified as the point where the Hessian matrix with (solid line) and without the stress term (dashed line) diverge. }
\end{figure}
\end {document}